\documentclass[preprint,12pt]{article}

\usepackage{amsmath,amssymb,array,calc,rotating,epsfig,psfrag, amscd}
\usepackage{color}
\usepackage[left=2cm,top=2cm,right=2cm,nohead]{geometry}

\usepackage[
      colorlinks=true,
      urlcolor=blue,    
      filecolor=blue,     
      citecolor=red,
      pdfstartview=FitV,
       bookmarksopen=true    
      ]{hyperref}

\definecolor{cardinal}{rgb}{0.6,0,0}
\definecolor{darkgreen}{rgb}{0,0.5,0}
\definecolor{golden}{rgb}{0.92, 0.7, 0}
\definecolor{midnight}{rgb}{0, 0, 0.5}
\definecolor{darkblue}{rgb}{0.2, 0, 0.8}

\newcommand{\nc}{\newcommand}

\nc{\ra}{\rightarrow} 
\nc{\lra}{\leftrightarrow} 
\nc{\Ra}{\Rightarrow} 
\nc{\LRa}{\Leftightarrow} 
\nc{\blp}{{\big (}}
\nc{\brp}{{\big )}}
\nc{\Blp}{{\Big (}}
\nc{\Brp}{{\Big )}}
\nc{\bglp}{{\bigg (}}
\nc{\bgrp}{{\bigg )}}
\nc{\Bglp}{{\Bigg (}}
\nc{\Bgrp}{{\Bigg )}}
\nc{\slb}{{\rm [}}
\nc{\srb}{{\rm ]}}
\nc{\bslb}{{\rm \big [}}
\nc{\bsrb}{{\rm \big ]}}
\nc{\Bslb}{{\rm \Big [}}
\nc{\Bsrb}{{\rm \Big ]}}
\def\al{\alpha}

\def\eps{\epsilon}
\nc{\veps}{\varepsilon}
\def\gam{\gamma}

\def\lam{\lambda}
\def\om{\omega}
\nc{\vphi}{\varphi}
\def\tha{\theta}
\def\sig{\sigma}

\def\Gam{\Gamma}

\def\Om{\Omega}
\def\Sig{\Sigma}

\nc{\myvspace}{\rule[-1em]{0pt}{2.5em}}
\nc{\bea}{\begin{eqnarray}}
\nc{\eea}{\end{eqnarray}}
\nc{\be}{\begin{equation}}
\nc{\ee}{\end{equation}}
\nc{\barr}{\begin{array}}
\nc{\earr}{\end{array}}
\nc{\cA}{{\cal A}}
\nc{\cB}{ \cal B}

\def\cD{{\cal D}}

\nc{\cF}{{\cal F}}
\nc{\cG}{{\cal G}}

\nc{\cL}{{\cal L}}
\nc{\cM}{{\cal M}}

\def\cO{{\cal O}}

\nc{\cQ}{{\cal Q}}
\nc{\cR}{{\cal R}}

\def\cV{{\cal V}}
\def\cV{{\cal V}}

\def\cZ{{\cal Z}}
\nc{\cQd}{\cQ^{\dagger}}
\nc{\cRd}{\cR^{\dagger}}
\nc{\BB}{{\mathbb B}}
\nc{\CC}{{\mathbb C}}
\nc{\DD}{{\mathbb D}}
\nc{\EE}{{\mathbb E}}
\nc{\FF}{{\mathbb F}}
\nc{\GG}{{\mathbb G}}
\nc{\HH}{{\mathbb H}}
\nc{\JJ}{{\mathbb J}}
\nc{\RR}{{\mathbb R}}
\nc{\PP}{{\mathbb P}}
\nc{\QQ}{{\mathbb Q}}
\nc{\ZZ}{{\mathbb Z}}
\nc{\calone}{{\mathbb 1}}

\nc{\half}{\frac{1}{2}}
\nc{\qrt}{\frac{1}{4}}
\nc{\del}{\partial}

\nc{\delbar}{\bar\partial}
\nc{\thalf}{\frac{t}{2}}
\nc{\Spin}{\operatorname{Spin}}
\nc{\SO}{\operatorname{SO}}

\nc{\Sp}{{\rm Sp}}
\nc{\com}[2]{{ \left[ #1, #2 \right] }}
\nc{\acom}[2]{{ \left\{ #1, #2 \right\} }}
\nc{\rr}{\rightarrow}
\nc{\p}{\partial}
\nc{\LT}{{\LL_\T}}
\nc{\Tr}{{\rm Tr}}
\nc{\tr}{{\rm tr}}
\def\com#1#2{{ \left[ #1, #2 \right] }}
\def\acom#1#2{{ \left\{ #1, #2 \right\} }}
\nc{\Adag}{A^{\dagger}}
\nc{\AdagI}{A^{\dagger I}}
\nc{\AdagJ}{A^{\dagger J}}
\nc{\AdagK}{A^{\dagger K}}
\nc{\AdagL}{A^{\dagger L}}
\nc{\AdagM}{A^{\dagger M}}
\nc{\Bdag}{B^{\dagger}}
\nc{\BdagI}{B^{\dagger}_I}
\nc{\BdagJ}{B^{\dagger}_J}
\nc{\BdagK}{B^{\dagger}_K}
\nc{\BdagL}{B^{\dagger}_L}
\nc{\BdagM}{B^{\dagger}_M}
\nc{\Cdag}{C^{\dagger}}
\nc{\CdagI}{C^{\dagger I}}
\nc{\CdagJ}{C^{\dagger J}}
\nc{\CdagK}{C^{\dagger K}}
\nc{\Ddag}{D^{\dagger}}
\nc{\DdagI}{D^{\dagger I}}
\nc{\DdagJ}{D^{\dagger J}}
\nc{\DdagK}{D^{\dagger K}}
\nc{\ttha}{\tilde{\theta}}
\nc{\tphi}{\tilde{\phi}}
\nc{\tsig}{\tilde{\sig}}
\nc{\tom}{\tilde{\om}}
\nc{\tlam}{\tilde{\lam}}
\nc{\tSig}{\widetilde{\Sig}}
\nc{\tPhi}{\tilde{\Phi}}
\nc{\tPhibar}{\ol{\tPhi}}
\nc{\tPi}{\tilde{\Pi}}
\nc{\tpsi}{\tilde{\psi}}
\nc{\tPsi}{\tilde{\Psi}}
\nc{\tgam}{\tilde{\gam}}
\nc{\tGam}{\tilde{\Gam}}
\nc{\tzeta}{\tilde{\zeta}}
\nc{\tZeta}{\tilde{\Zeta}}
\nc{\teta}{\tilde{\eta}}
\nc{\teps}{\tilde{\eps}}
\nc{\tEta}{\tilde{\Eta}}
\nc{\tchi}{\tilde{\chi}}
\nc{\tChi}{\tilde{\Chi}}
\nc{\txi}{\tilde{\xi}}
\nc{\tXi}{\tilde{\Xi}}
\nc{\tb}{\tilde b}
\nc{\tc}{\tilde c}
\nc{\te}{\tilde e}
\nc{\tf}{\tilde f}
\nc{\tg}{\tilde g}
\nc{\tj}{\tilde j}
\nc{\tp}{\widetilde{p}}
\nc{\tq}{\widetilde{q}}
\nc{\ts}{{\tilde s}}
\nc{\tu}{{\tilde u}}
\nc{\tv}{{\tilde v}}
\nc{\tw}{{\tilde w}}
\nc{\tx}{{\tilde x}}
\nc{\ty}{{\tilde y}}
\nc{\tz}{\tilde z}
\nc{\tA}{{\widetilde A}}
\nc{\tAbar}{{\ol \tA}}
\nc{\tB}{{\widetilde B}}
\nc{\tC}{{\widetilde C}}
\nc{\tD}{{\widetilde D}}
\nc{\tE}{{\widetilde E}}
\nc{\tG}{{\widetilde G}}
\nc{\tH}{{\widetilde H}}
\nc{\tJ}{{\widetilde J}}
\nc{\tJbar}{{\ol {\tilde J}}}
\nc{\tK}{{\widetilde K}}
\nc{\tL}{{\widetilde L}}
\nc{\tM}{{\widetilde M}}
\nc{\tN}{{\widetilde N}}
\nc{\tP}{{\widetilde P}}
\nc{\tQ}{{\widetilde Q}}
\nc{\tR}{{\widetilde R}}
\nc{\tS}{\widetilde{S}}
\nc{\tF}{\tilde{{\cal F}}}
\nc{\tX}{\widetilde{X}}
\nc{\tY}{\widetilde{Y}}
\nc{\tcZ}{\tilde{\cZ}}
\nc{\tcZbar}{\ol{\tcZ}}
\nc{\hb}{\hat b}
\nc{\hc}{\hat c}
\nc{\hd}{\hat d}
\nc{\he}{\hat e}
\nc{\hf}{\hat f}
\nc{\hg}{\hat g}
\nc{\hh}{\hat h}
\nc{\hp}{\hat p}
\nc{\hs}{\hat s}
\nc{\hv}{\hat v}
\nc{\hw}{\hat w}
\nc{\hx}{\hat x}
\nc{\hy}{\hat y}
\nc{\hz}{\hat z}
\nc{\zhat}{\hat z}
\nc{\hA}{\widehat{A}}
\nc{\hE}{\widehat{E}}
\nc{\hF}{\widehat{F}}
\nc{\hH}{\widehat{H}}
\nc{\hJ}{\widehat{J}}
\nc{\hK}{\widehat{K}}
\nc{\hL}{\widehat{L}}
\nc{\hM}{\widehat M}
\nc{\hN}{\widehat{N}}
\nc{\hV}{\widehat V}
\nc{\hcV}{\widehat \cV}
\nc{\hX}{\widehat X}

\nc{\ha}{\widehat \alpha}
\nc{\hphi}{\hat{\phi}}
\nc{\hpsi}{\hat{\psi}}
\nc{\hgam}{\hat{\gam}}
\nc{\hPhi}{\hat{\Phi}}
\nc{\hPsi}{\hat{\Psi}}
\nc{\hGam}{\hat{\Gam}}
\nc{\omhat}{\hat{\om}}
\nc{\hOm}{\widehat{\Om}}

\nc{\w}{\wedge}

\nc{\vb}{\vec b}
\nc{\vc}{\vec c}
\nc{\vd}{\vec d}
\nc{\ve}{\vec e}
\nc{\vf}{\vec f}
\nc{\vg}{\vec g}
\nc{\vh}{\vec h}
\nc{\vp}{\vec p}
\nc{\vq}{\vec q}
\nc{\vr}{\vec r}
\nc{\vs}{\vec s}
\nc{\vv}{\vec v}
\nc{\vw}{\vec w}
\nc{\vx}{\vec x}
\nc{\vy}{\vec y}
\nc{\vz}{\vec z}

\nc{\vB}{\vec B}
\nc{\vC}{\vec C}
\nc{\vD}{\vec D}
\nc{\vE}{\vec E}
\nc{\vF}{\vec F}
\nc{\vG}{\vec G}
\nc{\vH}{\vec H}
\nc{\vP}{\vec P}
\nc{\vQ}{\vec Q}
\nc{\vR}{\vec R}
\nc{\vS}{\vec S}
\nc{\vV}{\vec V}
\nc{\vW}{\vec W}
\nc{\vX}{\vec X}
\nc{\vY}{\vec Y}
\nc{\vZ}{\vec Z}

\nc{\ol}{\overline}
\nc{\abar}{\ol{a}}
\nc{\bbar}{\ol{b}}
\nc{\cbar}{\ol{c}}
\nc{\dbar}{\ol{d}}
\nc{\ebar}{\ol{e}}
\nc{\ibar}{\ol{\imath}}
\nc{\jbar}{\ol{\jmath}}
\nc{\kbar}{\ol{k}}
\nc{\lbar}{\ol{l}}
\nc{\mbar}{\ol{m}}
\nc{\nbar}{\ol{n}}
\nc{\pbar}{\ol{p}}
\nc{\qbar}{\ol{q}}
\nc{\ubar}{\ol{u}}
\nc{\vbar}{\ol{v}}
\nc{\wbar}{\ol{w}}
\nc{\xbar}{\ol{x}}
\nc{\ybar}{\ol{y}}
\nc{\zbar}{\ol{z}}

\nc{\Abar}{\ol{A}}
\nc{\Bbar}{\ol{B}}
\nc{\Cbar}{\ol{C}}
\nc{\Dbar}{\ol{D}}
\nc{\Ebar}{\ol{E}}
\nc{\Fbar}{\ol{F}}
\nc{\Jbar}{\ol{J}}
\nc{\Kbar}{\ol{K}}
\nc{\Lbar}{\ol{L}}
\nc{\Mbar}{\ol{M}}
\nc{\Nbar}{\ol{N}}
\nc{\Pbar}{\ol{P}}
\nc{\Qbar}{\ol{Q}}
\nc{\Rbar}{\ol{R}}
\nc{\Sbar}{\ol{S}}
\nc{\Tbar}{\ol{T}}
\nc{\Ubar}{\ol{U}}
\nc{\Vbar}{\ol{V}}
\nc{\Wbar}{\ol{W}}
\nc{\Xbar}{{\overline X}}
\nc{\Ybar}{{\overline Y}}
\nc{\Zbar}{{\overline Z}}
\nc{\cZbar}{{\overline \cZ}}

\nc{\epsbar}{\ol{\epsilon}}
\nc{\lambar}{\ol{\lambda}}
\nc{\zetabar}{\ol{\zeta}}
\nc{\Zetabar}{\ol{\Zeta}}
\nc{\psibar}{\ol{\psi}}
\nc{\Psibar}{\ol{\Psi}}
\nc{\phibar}{\ol{\phi}}
\nc{\Phibar}{\ol{\Phi}}
\nc{\chibar}{\ol{\chi}}
\nc{\mubar}{\ol{\mu}}
\nc{\nubar}{\ol{\nu}}
\nc{\rhobar}{\ol{\rho}}
\nc{\ombar}{\ol{\om}}
\nc{\Ombar}{\ol{\Om}}
\nc{\Deltabar}{\ol{\Delta}}
\nc{\Thetabar}{\ol{\Theta}}
\nc{\xibar}{\ol{\xi}}
\nc{\Xibar}{\ol{\Xi}}

\nc{\Dthbar}{\ol{\rm D3}}

\nc{\gdot}{\dot{g}}
\nc{\xdot}{\dot{x}}
\nc{\ydot}{\dot{y}}
\nc{\phidot}{\dot{\phi}}
\nc{\sinp}{s_{\phi}}
\nc{\cosp}{c_{\phi}}
\nc{\tanp}{t_{\phi}}
\nc{\spone}{s_{\phi_1}}
\nc{\cpone}{c_{\phi_1}}
\nc{\tpone}{t_{\phi_1}}
\nc{\sptwo}{s_{\phi_2}}
\nc{\cptwo}{c_{\phi_2}}
\nc{\tptwo}{t_{\phi_2}}
\nc{\spth}{s_{\phi_3}}
\nc{\cpth}{c_{\phi_3}}
\nc{\tpth}{t_{\phi_3}}
\nc{\calp}{c_{\al}}
\nc{\salp}{s_{\al}}

\nc{\csch}{{\rm csch}}
\nc{\sech}{{\rm sech}}

\nc{\cothzlami}{\coth(z-\lam_i)}
\nc{\coshzlami}{\cosh(z-\lam_i)}
\nc{\sinhzlami}{\sinh(z-\lam_i)}

\nc{\cothzlamj}{\coth(z-\lam_j)}
\nc{\coshzlamj}{\cosh(z-\lam_j)}
\nc{\sinhzlamj}{\sinh(z-\lam_j)}

\nc{\cothlamij}{\coth(\lam_i-\lam_j)}
\nc{\coshlamij}{\cosh(\lam_i-\lam_j)}
\nc{\sinhlamij}{\sinh(\lam_i-\lam_j)}
\nc{\bah}{{\mathbf {\hat{A}}}}
\nc{\bX}{{\mathbf X}}
\nc{\ba}{{\bf a}}
\nc{\bb}{{\bf b}}
\nc{\bc}{{\bf c}}
\nc{\bd}{{\bf d}}
\nc{\bg}{{\bf g}}
\nc{\bk}{{\bf k}}
\nc{\bl}{{\bf l}}
\nc{\bm}{{\bf m}}
\nc{\bn}{{\bf n}}
\nc{\bo}{{\bf o}}
\nc{\bp}{{\bf p}}
\nc{\bq}{{\bf q}}
\nc{\br}{{\bf r}}
\nc{\bs}{{\bf s}}
\nc{\bt}{{\bf t}}
\nc{\bu}{{\bf u}}
\nc{\bv}{{\bf v}}
\nc{\bw}{{\bf w}}
\nc{\bx}{{\bf x}}
\nc{\by}{{\bf y}}
\nc{\bz}{{\bf z}}
\nc{\bom}{{\bf \om}}
\nc{\bombar}{{\mathbf \ombar}}
\nc{\bPhi}{{\bf \Phi}}

\nc{\rma}{{\rm a}}
\nc{\rmb}{{\rm b}}
\nc{\rmc}{{\rm c}}
\nc{\rmd}{{\rm d}}
\nc{\rmg}{{\rm g}}
\nc{\rk}{{\rm k}}
\nc{\rml}{{\rm l}}
\nc{\rmm}{{\rm m}}
\nc{\rmn}{{\rm n}}
\nc{\rmo}{{\rm o}}
\nc{\rmp}{{\rm p}}
\nc{\rmq}{{\rm q}}
\nc{\rmr}{{\rm r}}
\nc{\rms}{{\rm s}}
\nc{\rmt}{{\rm t}}
\nc{\rmu}{{\rm u}}
\nc{\rmv}{{\rm v}}
\nc{\rmw}{{\rm w}}
\nc{\rmx}{{\rm x}}
\nc{\rmy}{{\rm y}}
\nc{\rmz}{{\rm z}}
\nc{\Ffour}{{F^{(4)}}}
\nc{\Ffive}{{F^{(5)}}}

\nc{\dal}{\dot{\al}}
\nc{\thadot}{\dot{\tha}}
\nc{\thab}{\bar{\theta}}
\nc{\thal}{\theta^{\al}}
\nc{\thdal}{\bar{\theta}^{\dal}}

\nc{\thsigthm}{\tha \sigma^m \thab}
\nc{\thsigthn}{\tha \sigma^n \thab}

\nc{\Dal}{D_{\al}}
\nc{\Ddal}{\bar{D}_{\dal}}
\nc{\CDal}{{\cal D}_{\al}}
\nc{\CDdal}{\bar{\cal D}_{\dal}}

\nc{\eq}[1]{(\ref{#1})}
\nc{\non}{\nonumber}

\nc{\equ}{{\rm eq}}

\nc{\vol}{{\rm vol}}
\nc{\Ainf}{A_{\infty}}
\nc{\End}{{\rm End}}
\nc{\Ext}{{\rm Ext}}
\nc{\IIB}{{\rm IIB}}
\nc{\Ad}{{\rm Ad}}
\nc{\IIA}{{\rm IIA}}
\nc{\AdS}{{\rm AdS}}
\nc{\CFT}{{\rm CFT}}
\nc{\Dslash}{\ensuremath \raisebox{0.025cm}{\slash}\hspace{-0.32cm} D}
\nc{\cDslash}{\ensuremath \raisebox{0.025cm}{\slash}\hspace{-0.32cm} \cD}
\nc{\no}{\!:\!\!}
\nc{\ointdz}{\oint\frac{dz}{2\pi i}}
\nc{\ointdzone}{\oint\frac{dz_1}{2\pi i}}
\nc{\ointdztwo}{\oint\frac{dz_2}{2\pi i}}
\nc{\ointdzb}{\oint\frac{d\zbar}{2\pi i}}
\nc{\ointdzbone}{\oint\frac{d\zbar_1}{2\pi i}}
\nc{\ointdzbtwo}{\oint\frac{d\zbar_2}{2\pi i}}
\nc{\dz}{\frac{dz}{2\pi i}}
\nc{\dzb}{\frac{d\zbar}{2\pi i}}
\nc{\bpm}{\begin{pmatrix}}
\nc{\epm}{\end{pmatrix}}
 \nc{\bitem}{\begin{itemize}}
 \nc{\eitem}{\end{itemize}}

\begin{document}

\begin{flushright}
IPhT-T10/174
\end{flushright}

\vspace{0.5cm}
\begin{center}
\baselineskip=13pt {\LARGE \bf{On The Inflaton Potential From \\ Antibranes in Warped Throats}\\}
 \vskip1.5cm 
 Iosif Bena$^{\dagger}$, Gregory Giecold$^{\dagger}$, Mariana Gra\~na$^{\dagger}$ and Nick Halmagyi$^{\dagger *}$\\ 
 \vskip0.5cm
$^{\dagger}$\textit{Institut de Physique Th\'eorique,\\
CEA Saclay, CNRS URA 2306,\\
F-91191 Gif-sur-Yvette, France}\\
\vskip0.8cm
$^{*}$\textit{Laboratoire de Physique Th\'eorique et Hautes Energies,\\
Universit\'e Pierre et Marie Curie, CNRS UMR 7589, \\
F-75252 Paris Cedex 05, France}\\
\vskip0.5cm
iosif.bena, gregory.giecold, mariana.grana@cea.fr\\
halmagyi@lpthe.jussieu.fr \\ 

\end{center}

\begin{abstract}
  We compute the force between a stack of smeared antibranes at the
  bottom of a warped throat and a stack of smeared branes at some
  distance up the throat, both for anti--D3 branes and for anti--M2
  branes.  We perform this calculation in two ways: first, by treating
  the antibranes as probes in the background sourced by the branes
  and second, by treating the branes as probes in the candidate
  background sourced by the antibranes.  These two very different
  calculations yield exactly the same expression for the force, for
  all values of the brane--antibrane separation. This indicates that
the force between a brane and an antibrane is not screened in backgrounds where there is
  positive charge dissolved in flux, and gives a way to precisely
  compute the inflaton potential in certain string cosmology
  scenarios.

  \end{abstract}

\newpage

\section{Introduction and Motivation}

Anti--D3-branes in warped deformed conifold throats are widely used in
string theory model building and string cosmology, both to get de Sitter solutions
\cite{Kachru:2003aw}, and to construct string theoretic models of
inflation using D3 branes moving in such throats \cite{Kachru:2003sx}.

In a previous work~\cite{Bena:2009xk}, three of us
attempted to construct the first--order backreacted supergravity
solution for a stack of anti--D3 branes in the Klebanov--Strassler (KS)
background \cite{Klebanov:2000hb}. Such antibranes were conjectured in
\cite{Kachru:2002gs} to give rise to holographic duals to metastable
vacua of a strongly--coupled gauge theory, and the supergravity
analysis implies that the would--be anti--D3 brane solution must have a
certain infrared singularity.  A similar result was obtained by
investigating anti--M2 branes in a warped Stenzel
background~\cite{Bena:2010gs}. If these singularities have a physical
origin, then the solutions found in~\cite{Bena:2009xk, Bena:2010gs}
describe the first--order backreaction of antibranes in these
backgrounds. If these singularities are pathological, the analyses
of~\cite{Bena:2009xk, Bena:2010gs} imply that antibranes in
backgrounds with positive brane charge dissolved in fluxes cannot be treated in
perturbation theory.

In the present work we will work under the assumption that the
singularities found in~\cite{Bena:2009xk, Bena:2010gs} are physical,
and that antibranes can be treated as perturbations of their respective backgrounds
with charge dissolved in fluxes\footnote{Note that this does not
  automatically imply that antibranes give rise to metastable vacua --
  for this one would have to show also that the antibrane solution does
  not contain other non--normalizable modes.}.

In certain string inflation models, the inflaton is the position of a BPS D3 brane in a warped background with 
anti--D3 branes at its bottom, and the brane--antibrane force gives the derivative of the
inflaton potential. There exist two methods to compute this potential. 
The first, introduced in~\cite{Kachru:2003sx} and widely used
in string cosmology constructions, treats the anti--D3 branes as
probes in the (easy to find) backreacted solution sourced by BPS D3
branes up the throat. This method involves calculating the change in
the potential of the anti--D3 branes as the position of the D3 branes
is altered. This yields the force felt by these D3 branes in the
warped deformed conifold with anti--D3 branes.

The second method to derive the inflaton potential consists in
constructing the first--order backreacted solution sourced by anti--D3
branes placed at the bottom of a warped deformed
conifold~\cite{Bena:2009xk} and to compute the force felt by a probe
D3 brane in this background. Despite the rather complicated nature of
the first--order deformation space, the force on a probe D3 turns out
to depend only on one of the fourteen integration constants that
parametrize the space of $SU(2) \times SU(2) \times \ZZ_2$--invariant
deformations~\cite{Bena:2009xk}. Furthermore, the leading
large--distance behavior of the inflaton potential agrees with the one
computed in~\cite{Kachru:2003sx}.

One natural question to ask is whether the two calculations for the
inflaton potential agree also beyond leading--order. At first
glance, one expects that they should indeed agree, as this ought to be merely a
consequence of Newton's third law: the force exerted by the brane on
the antibrane is the same as the force exerted by the antibrane on the
brane~\cite{Kachru:2003sx}. 

However, the answer does not appear to be so simple. If in the vacuum
the calculations of the force using the bare action of one brane in the
background of the other should indeed agree, there is no reason this
should happen in a background where the charge/anticharge symmetry is
broken by the D3 charge dissolved in flux. Indeed, because of harmonic
superposition, the fields of the D3 brane are not
screened~\cite{Grana:2000jj}. Yet, there is no reason why the anti--D3
would not be screened by the D3 charge dissolved in flux. Hence, one
would expect to have a screening cloud around the anti--D3 branes,
which would affect the potential felt by a bare D3 brane. Note that
this is a generic phenomenon in media where positive and negative
charges are screened differently: because of the different profiles of
the screening clouds, the force computed using the action of a bare
negative charge in the background of the screened positive charge needs
not agree with the force computed using the action of a bare positive
charge around the screened negative charge.  In the language of plasma
physics, the Debye screening lengths of the positive and of the
negative charges need not be equal.

The purpose of this letter is to show that the forces computed in the
two approaches outlined above agree not only in leading behavior, but
in full functional form, modulo a to--be--determined overall
normalization constant. This indicates that this force is not
screened by the brane charge dissolved in flux\footnote{Our analysis does not formally exclude
screening by a delta--function--shaped screening cloud, which would keep the same functional expression 
of the force while changing the overall normalization constant. However, it is hard to believe this is 
anything but a formal possibility. We leave the actual
computation of this constant to a forthcoming
publication~\cite{wip}.}. There are two obvious explanations for this: either anti-D3 branes are
not screened by the positive D3 brane charge dissolved in flux, or
they are screened, but the screening cloud does not interact with D3 branes. This latter possibility would 
imply that antibranes change the profile of the cloud of charge 
dissolved in fluxes, but do not alter its properties, in particular the fact that the local D3 charge density 
remains equal to the mass density; such a cloud would not interact with probe D3 branes and would not screen the force.

We find no brane--antibrane force screening, both for anti D3--branes at the bottom of the
Klebanov--Strassler solution, and for anti--M2 brane at the bottom of
a warped Stenzel space with M2 brane charge dissolved in
flux~\cite{Stenzel:1993, Cvetic:2000db}, and hence we believe this is likely a
generic phenomenon in flux compactifications.\footnote{In an upcoming paper~\cite{Giecold:D2} we will also show
  this for anti--D2 branes in backgrounds with D2 brane charge
  dissolved in fluxes~\cite{Cvetic:2001ma}.} 
  
Hence, in an optimistic scenario (if the IR singularities found in~\cite{Bena:2009xk} 
and~\cite{Bena:2010gs} are physical, and we can trust perturbation theory), modulo this subtle issue about the
overall constant, our calculation yields the exact functional form of
the inflaton potential in a brane/antibrane realization of
inflation in string theory. It also demonstrates that the force between branes and antibranes is
not screened, and therefore the probe antibrane calculation \` a la
KKLMMT~\cite{Kachru:2003sx} of this inflaton potential in other string
inflationary models gives the exact functional form of the potential,
not only its leading behavior.  This should allow in turn to
accurately compute the power spectrum in those models and to compare
them with observation.

The paper is organized as follows. In Section 2 we review the calculation of the
brane/antibrane force, treating the smeared antibranes as probes, both for anti--D3 branes 
in KS and for anti--M2 branes in a warped Stenzel background~\cite{Stenzel:1993, Cvetic:2000db}. In Section 3 we use the
first--order backreacted solutions of~\cite{Bena:2009xk} and~\cite{Bena:2010gs} to
compute this force using the action of probe D3 and M2
branes, respectively. As advertised, the two calculations agree.

\noindent {\bf Note:} as this paper was being prepared for submission we learnt 
that Anatoly Dymarsky has independently found some of the analytic results 
presented here.

\section{Computing the Force Using the Action of Probe Antibranes}

To establish whether antibranes are screened by charge dissolved in
flux in the warped deformed conifold or the Stenzel space, we first
smear them at the tip of those two geometries. This way we preserve
the symmetries of the solution without antibranes, and render the
calculation of the backreaction of the antibranes an achievable task.
The force between the smeared antibranes and the BPS branes at some
distance $r = r_0$ up the throat will then be the same, by symmetry,
as the force between the smeared antibranes and a uniform shell of BPS
branes at the same distance.

We demonstrate how to compute the force generated between a stack of
antibranes at the bottom of a warped throat and a stack of branes some
distance up the throat. This is computed in two ways: either by
backreacting the branes while leaving the antibranes as probes ; or
from backreacting the antibranes and leaving the branes as probes.

\subsection{Backreacted D3 Branes in the Warped Deformed Conifold} \label{sec:GreenConifold}

To obtain a fully backreacted solution with 
BPS D3 branes in the warped deformed conifold one simply needs to add to the warp factor a harmonic 
function (given by the Green's function on this Calabi-Yau manifold) sourced by these branes~\cite{Grana:2000jj}.
 While in
general this is a non--trivial task~\cite{Krishnan:2008gx,
  Pufu:2010ie}, here we are considering smeared branes and as such the
Green's function is radially symmetric and the problem is tractable.

The two radially symmetric solutions to the Laplace equation on the deformed conifold are
\bea
H_1(\tau) &=& c_1 \, , \\
H_2(\tau)&=& c_2\, \int_\tau^{\infty} \frac{d\tau'}{ \blp \sinh 2\, \tau' - 2\, \tau' \brp^{2/3}} \, .
\eea
With a shell of D3 branes at $\tau=\tau_0$, the full warp factor is
\be
H(\tau) = H_{0}(\tau) + \delta H(\tau) \ .
\ee
Here $H_{0}(\tau)$ is the zeroth--order warp factor for the warped deformed conifold:
\bea
 H_{0}&=&e^{-4A_0-4p_0+2x_0} \non \\
 &=&h_0 - 32\, P^2\, \int_0^\tau \frac{t\, \coth t -1}{\sinh^2t} (\tfrac12 \, \sinh(2\, t) - t)^{1/3}\, dt \ , \label{KSh}
 \eea
where $P$ is the RR three--form flux through the $S^3$ of the deformed conifold, and $h_0$ is a constant\footnote{Explicitly, we have $h_0=32\, P^2\, \int_0^{\infty} \frac{\tau\, \coth \tau -1}{\sinh^2 \tau} (\tfrac12 \, \sinh(2 \, \tau) - \tau)^{1/3}\, d\tau=18.2373 P^2$.}. 
On top of the warp factor for the zeroth--order solution, there is the following contribution from the $N$ D3 branes at $\tau = \tau_0$ :
\be
\delta H(\tau) = {\Bigg \{} \begin{array}{cl}
H_{1}(\tau) \, , & \tau<\tau_0 \, , \\
H_{2}(\tau) \, , & \tau> \tau_0 \ .
\end{array}
\ee
The two integration constants $(c_1, c_2)$ are related by matching at the source the solutions in the two domains above:
\be
c_1=H_{2}(\tau_0) \, .
\ee
To fix the other integration constant in terms of the number of D3 branes, we rely on the standard quantization formula for the five--form field strength:
\be
\frac{1}{(4\, \pi^2\, \alpha')^2} \int F_5 = N \, ,
\ee
and integrate on the $T^{1,1}$ surfaces right outside and right inside the shell using 
\be
g_s F_5=*_{10}d H^{-1}\w dx_0\w\ldots\w dx_3 \, .
\ee
The difference of the two integrals gives the D3 brane charge of the shell and its relation to the coefficient in $\delta H$:
\be\label{c2 in term of N}
c_2 = 4\, \pi\, \Blp\frac{2^{1/3}\, \al'}{\veps^{4/3}}\Brp^2 \, g_s \,N \, ,
\ee
where we use the conventions of~\cite{Herzog:2002ih}.

We now compute the potential of
probe anti--D3 branes placed at the tip of the cone. Since for a
BPS D3 brane the DBI and WZ potentials cancel, for anti--D3 branes these potentials are equal in magnitude and sign:

\be
V_{\ol{D3}}= V_{DBI}+ V_{WZ}= 2\, V_{WZ}  \, .
\ee
Expanding the potential to first--order in the number of D3 branes we find 
\bea
V_{\ol{D3}  }&=& 2\, H^{-1} \, , \non \\
&=& 2\, H_{0}^{-1}(1-\frac{\delta H}{H_{0}}) +\cO((N/P)^2).
\eea

The force exerted by the anti--D3 branes on the D3 branes can then be obtained from the variation of this potential as the source D3 branes are moved~\cite{Kachru:2003sx}
\bea
F_{D3}&=& - \frac{\del V_{\ol{D3}}}{\del \tau_0} {\Big |}_{\tau=0}\non \\
&=&-\frac{1}{H^2_{0}|_{\tau=0}}\,  \frac{c_2}{(\sinh 2\, \tau_0-2\, \tau_0)^{2/3} }\label{probeantiD3force} \, .
\eea
The dependence of this force on $N$ appears through the constant $c_2$~\eqref{c2 in term of N}.

\subsection{M--Theory on a Warped Stenzel Space}    \label{sec:GreenStenzel}

The generalization of the probe brane computation of Kachru, Pearson
and Verlinde~\cite{Kachru:2002gs} to a warped Stenzel
space M--theory background~\cite{Stenzel:1993, Cvetic:2000db} has recently been
performed in~\cite{Klebanov:2010qs}. Motivated by this analysis, three of the authors have used the technology of~\cite{Bena:2009xk} to study the backreaction of anti--M2 branes in this space~\cite{Bena:2010gs}. The probe
brane analysis of the previous section can also be performed, and we
find that although the Green's function itself is a complicated combination of incomplete elliptic integrals:
\begin{align}
H_{1}(y) = &d_1 \, , \nonumber \\
H_{2}(y) =&\, \frac{2}{45}\, d_2\, \Big[ \frac{9\, \sqrt{y^4-1}}{y^5} + 3\, E\left(\arcsin(1/y)\mid-1\right) - 3\, F\left(\arcsin(1/y)\mid-1\right) \nonumber\\ & + 5\, \sqrt{3}\, \left( \Pi\left(\sqrt{3}; - \arcsin(1/y)\mid-1\right) - \Pi\left(-\sqrt{3}; -\arcsin(1/y)\mid-1\right)\right)\Big] \, ,
\end{align}
with $d_i$ integration constants and $k$ a constant that ensures that $H_2$ vanishes at large $y$, the derivative of this Green's function is very simple\footnote{The standard coordinate we use is $y^4=2+\cosh2\, r$.}:
\be
H'_{2}(r)=\frac{3\, \sqrt{2}\, d_2\, \csch^3 r}{(2+\cosh 2\, r )^{3/4}} \, .
\ee
From flux quantization 
\be
\frac{1}{(2\, \pi\, \ell_p)^6}\, \int_{V_{5,2}} *_{11} G_{4} = N \, ,
\ee
with
\be
G_4= d H^{-1} \w dx_0\w dx_1\w dx_2 \, , \non \\
\ee
we find that the M2 brane charge of the shell, $N$, is related to the constant in the new warp factor via
\be
d_2=(2\, \pi)^2\, \ell_p^6\, \sqrt{2}\, N \, .
\ee
In addition, there is the matching condition
\be
d_1=H_2(y_0) \, .
\ee

If we now consider the change in the potential of probe antibranes with the position of the source M2 branes in this background, we obtain the force:
\be
F_{M2}= - \frac{1}{H^2_{0}|_{r=0}} \frac{ 3\, \sqrt{2}\, d_2\, \csch^3 r_0}{(2+\cosh 2\, r_0 )^{3/4}} \, . \label{probeantiM2force}
\ee

\section{Computing the Force on Probe Branes}

\subsection{Warped Deformed Conifold}

We now use the results from~\cite{Bena:2009xk} and refer to this work
for much of the notation. In that paper three of the authors found
that the force felt by a probe D3 brane in the first--order deformed KS
background has the remarkably--simple form
\be
F_{D3}=\frac{2}{3}\, e^{-2x_0}\, \txi_1 \, ,
\ee
where $\txi_1$ is one of the sixteen modes parameterizing the
deformation space\footnote{This deformation space has been considered
  previously in various respects~\cite{Kuperstein:2003yt,Berg:2005pd,Berg:2006xy, Benna:2007mb,Dymarsky:2008wd}.} ~\cite{Borokhov:2002fm} and is given by 
\be
\txi_1=\tX_1\, \exp \Blp \int_0^\tau d\tau' e^{-2x_0}\bslb 2\, P\, f_0 - F_0\, (f_0 - k_0) \bsrb \Brp \, .
\ee
Here $X_1$ is an integration constant and
\bea \label{KSbackground}
 e^{x_0}&=& \frac14 \, H_0^{1/2}\, (\tfrac12 \, \sinh(2 \, \tau) - \tau)^{1/3} \, , \nonumber \\
 f_0&=&-P\, \frac{(\tau \, \coth \tau -1)(\cosh \tau -1)}{\sinh \tau} \, , \\
 k_0&=&-P\, \frac{(\tau \, \coth \tau -1)(\cosh \tau +1)}{\sinh \tau} \, , \nonumber \\
  F_0&=& P\, \frac{(\sinh \tau -\tau)}{\sinh \tau} \, , \nonumber 
 \eea
with $H_0$ given in (\ref{KSh}).

We make great use of the simple yet elusive observation that this integral can in fact be performed exactly
\bea
\txi_1&=&\tX_1\, \exp \Blp \int^\tau_0d\tau'\, \frac{H_0'}{H_0}  \Brp \nonumber \\
&=& X_1\, H_0(\tau) \, . \label{txi1}
\eea
The force now takes the form
\bea
F_{D3}&=&\frac{2}{3}\, e^{-2x_0}\, X_1\, H_0(\tau) \non \\
&=&\frac{32}{3}\, \frac{2^{2/3}\, X_1}{(\sinh 2\, \tau - 2\, \tau)^{2/3}}\label{probeD3force} \, .
\eea

Remarkably enough, this has exactly the same functional form as the
force computed in (\ref{probeantiD3force}) using the probe antibrane
potential. As mentioned in the Introduction, the fact that the two
calculations of the force agree implies that this force is not 
screened by the positive D3 brane charge dissolved in flux. 

  As has been explained in~\cite{Bena:2009xk} the value of $X_1$ can
  be determined in terms of the UV and IR boundary conditions, but this requires
  relating the UV and IR values of all sixteen integration constants involved in the full solution,
  which can only be done numerically. Once this numerical work is
  completed, we will be able to compare the coefficient of the force
  computed in this section with the calculation of section
  \ref{sec:GreenConifold}. Whether these two numbers agree or not will
  help elucidate the physics of anti--D3 branes in the
  Klebanov--Strassler background. We plan to report on these results
  soon~\cite{wip}.

\subsection{M--Theory on a Warped Stenzel Space}

The same steps for M--theory on a Stenzel space have recently been performed in~\cite{Bena:2010gs} and we merely quote the results and refer to this work for the notation.
When considering the candidate backreacted solution corresponding to 
anti--M2 branes, the force felt by  a probe M2 brane is 
\bea
F& = &- \frac{2}{3}\, e^{-3\,(\alpha_0 + \beta_0)(r)}\, e^{-3 z_0(0)}\, X_4  \nonumber\\
& =& - \frac{18\, e^{-3 z_0(0)}\, X_4 \, \csch^3 r}{\left( 2 + \cosh2 r \right)^{3/4}}\,  \, .\label{probeM2force}
\eea
This has again the same functional form as \eq{probeantiM2force}, up
to the determination of the integration constant $X_4$ in terms of the
charges of the system. This demonstrates that, much like in the
anti--D3 brane story, the force between anti--M2 branes and M2 branes is not screened by the charge dissolved in flux.


\vspace{1cm}
\noindent {\bf Acknowledgements}:
\noindent We would like thank Jean--Paul Blaizot and Jean--Yves Ollitrault for informative discussions. The work of G.~G.~is supported by a Contrat de Formation par la Recherche of CEA/Saclay. The work of I.~B., M.~G.~and N.~H.~is supported by the DSM CEA/Saclay, the grant ANR--07--CEXC--006, the ANR grant 08--JCJC--0001--0, and by the ERC Starting Independent Researcher Grant 240210 -- String--QCD--BH.

\providecommand{\href}[2]{#2}\begingroup\raggedright\endgroup

\end{document}